\documentclass[11pt]{article} \usepackage{moriond,epsfig}
\def\PLB{{\em Phys. Lett.}  B}
\def\PRD{{\em Phys. Rev.} D}
\newcommand {\EPJ}    {{\em Eur.{} Phys.{} J.{} }}

\def\ra{\ensuremath{\rightarrow}}
\def\stat{\mbox{$\;$(stat.)}}
\def\syst{\mbox{$\;$(syst.)}}
\def\lumi{\mbox{$\;$(lumi.)}}
\def\met{\ensuremath{E_{\mathrm{T}}^{\mathrm{miss}}}}
\def\ttbar{\ensuremath{t\bar{t}}}
\def\stt{\ensuremath{\sigma_{t\bar{t}}}}
\def\st{\ensuremath{\sigma_{t}}}
\def\mtop{\ensuremath{m_{\mathrm{top}}}}
\def\pt{\ensuremath{p_{\mathrm{T}}}}
\def\TeV{\ifmmode {\mathrm{\ Te\kern -0.1em V}}\else
                   \textrm{Te\kern -0.1em V}\fi}%
\def\GeV{\ifmmode {\mathrm{\ Ge\kern -0.1em V}}\else
                   \textrm{Ge\kern -0.1em V}\fi}%
\def\ipb{\mbox{pb$^{-1}$}}%  Inverse picobarns.
\def\bfig{\begin{figure}[htbp]\centering}
\def\efig#1#2{\caption{#1}\label{fig:#2}\end{figure}}
\def\igra#1#2{\includegraphics[height=#2\textheight]{#1}}
\def\twoplot#1#2#3#4#5#6{\bfig\igra{#3}{#4}\igra{#5}{#6}\efig{#2}{#1}}

\def\conf#1#2{\bibitem{c#1}ATLAS Collaboration, ATLAS-CONF-2011-0#1, http://cdsweb.cern.ch/record/#2}

\begin{document}
\vspace*{4cm}
\title{RECENT RESULTS ON TOP PHYSICS AT ATLAS}
\author{M. CRISTINZIANI\\{\em on behalf of the ATLAS Collaboration\\[1ex]}}
\address{Physikalisches Institut, Universit\"{a}t Bonn,
Nussallee 12, 53127 Bonn, Germany}
\maketitle\abstracts{
During the 2010 $pp$ run of the Large Hadron Collider at $\sqrt{s} = 7 \TeV$, 
a substantial data sample of high $\pt$ triggers, 35 \ipb, has been collected
by the ATLAS detector, corresponding to about 2,500 produced top-quark 
pair events
containing at least one lepton ($e$ or $\mu$) in the final state. 
Measurements of the top-quark
pair production cross-section, the top mass, the $W$ helicity fractions in
top-quark decays and studies of single-top quark production and
top-quark pair production with anomalous missing transverse energy are presented.
}

\section{Introduction}
Top-quark measurements are of central importance to the LHC physics programme.
The production of top-quark pairs in 
$pp$ collisions is a process which is situated at the boundary between the 
Standard Model (SM) and what might lie beyond it.
Within the SM, top quarks are predicted to almost always 
decay to a $W$-boson and a $b$-quark. The decay topologies are determined
by the decays of the $W$-boson. In pair-produced top-quarks the single-lepton
and dilepton modes, with branching ratios of $37.9\%$ and $6.5\%$ respectively,
give rise to final states with one or two leptons (electrons or muons), missing transverse energy (\met)
and jets, some with $b$-flavour.

\section{Top quark pair production cross-section $\stt$}
The measurement of $\stt$ is a milestone for early LHC physics.
Within the SM the $\ttbar$ production cross-section at $\sqrt{s} = 7 \TeV$
is calculated to be $165^{+11}_{-16}$ pb at approximate NNLO~\cite{aNNLO} for 
a top-quark mass of $172.5 \GeV$. 
A precise determination of 
$\stt$ tests these perturbative QCD predictions.
First measurements of $\stt$ at the LHC
have been reported by ATLAS~\cite{atpap} and CMS~\cite{cmspap} with 3 \ipb.
Here, approximately ten times more data have been analysed.
\subsection{Dilepton channel}
The cross-section in the dilepton channel is extracted with a cut-and-count
method~\cite{c34}. Candidate events are selected by requiring two 
opposite-signed high-$\pt$ leptons in the $ee$, $\mu\mu$ and $e\mu$ topologies, 
and at least two jets. The background contribution from 
Drell-Yan production is suppressed by requiring 
for same-flavour events large $\met$ and 
for $e\mu$ events large $H_T$, 
the scalar sum of jet and lepton transverse energies.
Remaining Drell-Yan events and
background from fake leptons are estimated with data-driven methods.
Across the three channels 105 events are selected with an expected S/B ratio 
of 3.6.
The cross-section is extracted with a profile likelihood 
technique, with a simultaneous fit to the three channels and taking into
account systematic uncertainties. This results in 
$\stt = 174 \pm 23 \stat^{+19}_{-17} \syst \pm 7 \lumi$ pb.

Additional studies are performed to corroborate this measurement: 
a technique that normalizes the $\ttbar$ signal yield to the measured 
rate of $Z$-boson decays; a two-dimensional template shape fit using 
the $\met$ vs $N_{\mathrm{jets}}$ variables to simultaneously measure 
the production cross-sections of the $\ttbar$, $WW$ and $Z \ra \tau\tau$ 
final states; and a cross-section measurement that requires at least one 
$b$-tagged jet and a looser kinematic selection to optimize the S/B ratio.
All measurements are in good agreement with each other.
\subsection{Single-lepton channel}
Two complementary measurements have been performed in the single-lepton channel.
In the first measurement no explicit identification of secondary vertices inside jets ($b$-tagging) is
performed~\cite{c23}. 
The main background consists of $W$+jet events and QCD multi-jet events, where one jet mimics a reconstructed lepton.
The latter are particularly difficult to simulate correctly and are thus estimated using data-driven techniques.
Selected events are classified according to lepton flavour (2009 candidate 
events observed in $\mu$+jets and 1181 in $e$+jets) and according to
jet multiplicity: exactly 3 jets or $\ge 4$ jets. 
Three variables that exploit the different kinematic behaviour of
$\ttbar$ and the $W$+jets background events are identified and used in a multivariate
likelihood fit to extract $\stt = 171 \pm 17 \stat^{+20}_{-17} \syst \pm 6 \lumi$ pb.
The main systematic uncertainties are due to the limited knowledge of the jet energy scale and 
reconstruction efficiency as well as the amount of initial and final state radiation.

A second method exploits the $b$-tagging capabilities, albeit making use of a 
simple and robust tagging algorithm with a modest rejection factor~\cite{c35}. A multivariate 
likelihood discriminant is constructed using template distributions of four variables, 
among which the average of the weights of the most significant $b$-tags. Here, data are 
further split with an additional jet-bin (3, 4 or $\ge 5$ jets). A profile likelihood
technique is used to extract $\stt$ and constrain the systematic effects from data.
The result is $\stt = 186 \pm 10 \stat^{+21}_{-20} \syst \pm 6 \lumi \mathrm{pb}$, where the
systematic uncertainties are dominated by the uncertainties in the $b$-tagging algorithm
calibration from data and the heavy flavour fraction in $W$+jets events.
Cross-check measurements are performed with kinematic fits of 
the reconstructed top mass and cut-and-count methods and are 
found to be in good agreement with this result. 

\subsection{Combination}
The most precise cross-section measurements in the dilepton and single-lepton channels
are combined, taking into account correlated systematic uncertainties \cite{c40}. 
The result has a total uncertainty of 10\%, $\stt = 180 \pm 9 \stat \pm 15 \syst \pm 6 \lumi$ pb, 
and is in excellent agreement with the SM prediction as shown in Figure~1 (right).
\twoplot{figure}{Selected dilepton events superimposed on expectations from simulation and data-driven estimations (left). Cross-section measurements at Tevatron and LHC
compared to approximate NNLO predictions (right).}
{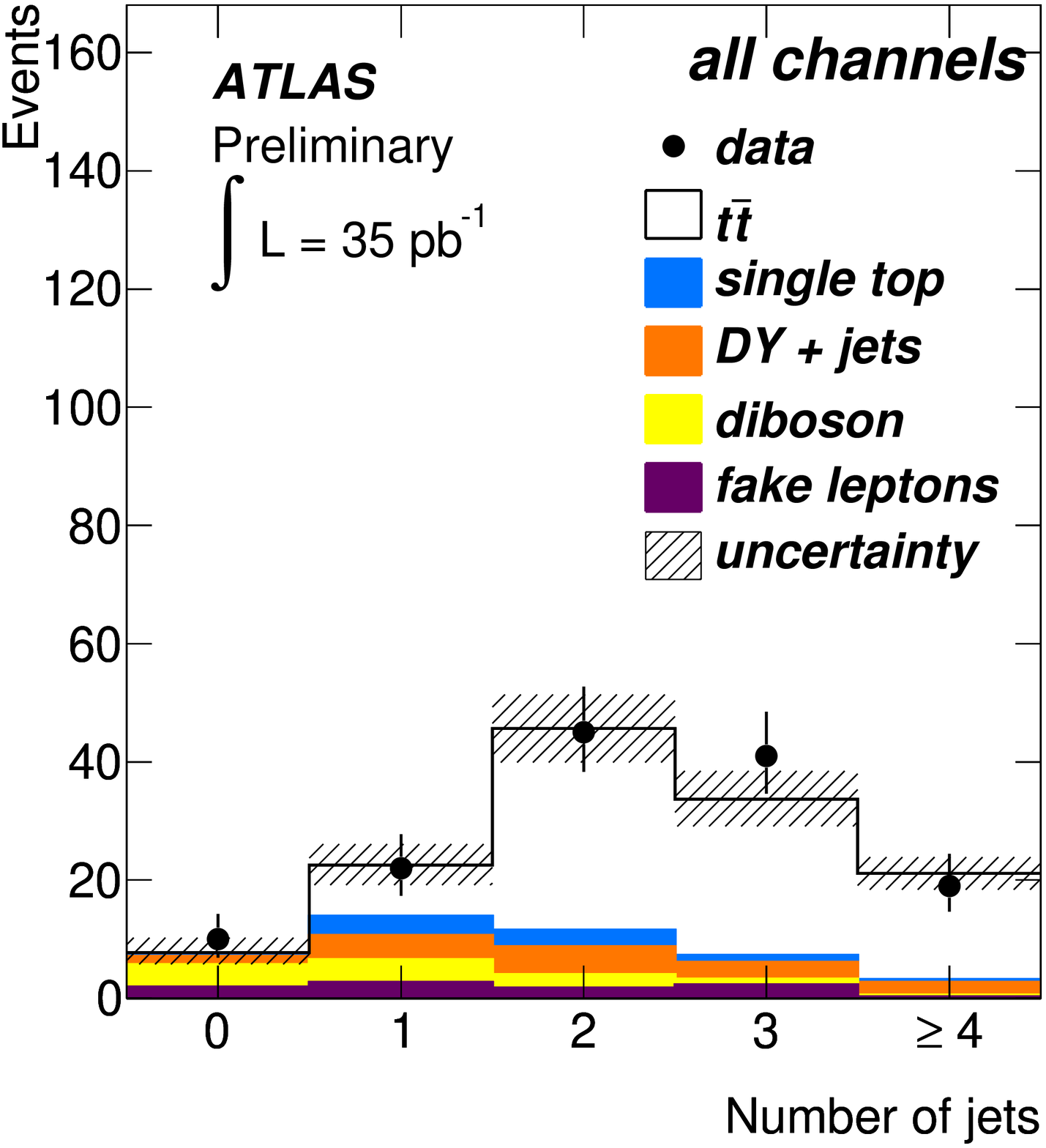}{0.23}{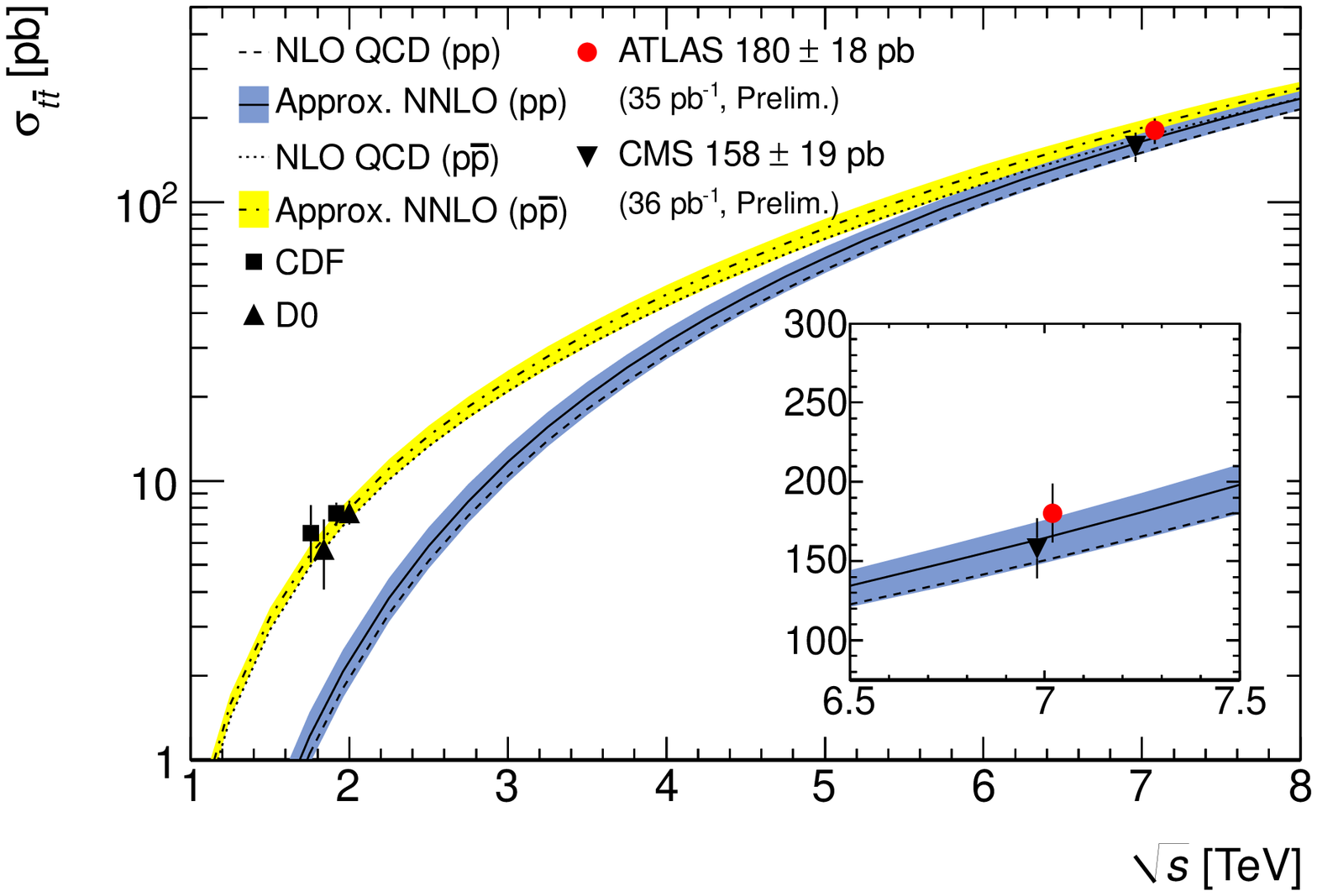}{0.23}

\section{Single-top production}
The observation of electroweak
production of single-top quarks has been reported by the CDF and D0 collaborations
in 2009. This final state provides a direct probe of the $W$-$t$-$b$ coupling and is 
sensitive to many models of new physics. The measurement of the production 
cross-section determines the magnitude of the quark mixing matrix element $V_{tb}$
without assumptions on the number of quark generations.
With the available data sample searches for single-top quark production in 
the $t$- and $Wt$-channels are performed~\cite{c27}. 

The $t$-channel search is based on the selection of events with exactly one 
identified lepton, jets and $\met$.
In a cut-based analysis a reconstructed three-jet invariant mass compatible with $\mtop$
is required, as well as the leading jet to be in the forward direction.
This selects 32 candidate events.
Using data-driven methods to estimate the QCD and $W$+jets 
backgrounds, a production cross-section of  $\st = 53^{+46}_{-36}$ pb is measured,
which translates to an upper limit of 162 pb at 95\% confidence level. A likelihood function approach is also 
used to cross-check the result. Both results are consistent with the 
SM expectation of 66 pb. 

The $Wt$-channel analysis is based on the selection of events with either one or two leptons, 
jets and $\met$. The expected SM cross-section for this 
single-top process is 15 pb. A 95\% confidence level limit 
is set on the $Wt$-channel production cross section of $\sigma_{Wt} < 158$ pb.
In the dilepton channel, the $\ttbar$ background is estimated from data, by considering
the one-jet bin as a control region.

\section{Top quark properties}
\subsection{Mass}
The top-quark mass, a fundamental parameter of the SM, is a source of large contributions
to electroweak radiative corrections and, in conjunction with precision electroweak
measurements, can be used to derive constraints on the Higgs boson mass or heavy 
particles predicted in SM extensions. The current world average is $\mtop = 173.3 \pm 1.1 \GeV$.

The main systematic on the determination of the top-quark mass is the uncertainty in the jet energy scale (JES).
Three complementary template analyses have been developed~\cite{c33}
that address the uncertainty due to the JES in different ways:
a 2D analysis that simultaneously determines $\mtop$ 
and a global jet energy scale factor between data and predictions;
a 1D analysis exploiting a kinematical likelihood 
fit to all decay products of the $\ttbar$ system;
a 1D analysis which is based on 
the ratio between the per-event reconstructed invariant
mass of the top-quark and the mass of the $W$-boson associated to the 
hadronically decaying top-quark candidate.
The latter method yields a top-quark mass measurement of 
\mbox{$\mtop = 169.3 \pm 4.0 \pm 4.9 \GeV$}.
\subsection{$W$-boson polarisation in top-quark decays}
The polarisation states of the $W$-bosons that emerge from top-quark decays are 
well defined in the SM, due to the $V-A$ structure of the charged current weak interactions.
These can be extracted from the angular distributions of the decay products in 
$t \ra bW \ra b\ell \nu_{\ell}$.

In a first measurement~\cite{c37}, templates of the $\cos \theta^{\star}$ distribution are built 
from simulation and fitted to selected events with a single charged lepton, 
\met\ and at least four jets, where at least one of them is $b$-tagged. 
Here $\theta^{\star}$ is the angle between the 
direction of the lepton and the reversed momentum direction of the $b$ quark
from the top-quark decay, both boosted into the $W$-boson rest frame. Events are 
reconstructed in the single-lepton channel with a kinematic fit method. 
Assuming $F_R = 0$
helicity fractions $F_0 = 0.59 \pm 0.12$ and $F_L = 0.41 \pm 0.12$ are extracted.
A second measurement is based on angular asymmetries constructed from the $\cos \theta^{\star}$
variable, the events are reconstructed with a $\chi^2$ fitting technique and an iterative
procedure is applied to correct for detector and reconstruction effects in order to recover
the undistorted distribution on parton level. The helicity fractions are measured to 
be $F_0 = 0.65 \pm 0.15$, $F_L = 0.36 \pm 0.10$ and $F_R = -0.01 \pm 0.07$.
Both results are in good agreement with the SM prediction and are used to place limits
on anomalous couplings $V_R$, $g_L$ and $g_R$ that arise in new physics models.
\subsection{Search for anomalous $\met$ in $\ttbar$ events}
A search for anomalous $\met$ in the single-lepton
$\ttbar$ final state has been performed~\cite{c36}. Such a phenomenon can arise from a number of 
extensions of the SM, but the focus here is on a search for a 
pair-produced exotic top partner $T$ with mass $m(T)$, that decays to a top-quark and a 
long lived neutral scalar particle $A_0$.
The final state for such a model is identical to $\ttbar$, but with a large amount of $\met$
from the undetected $A_0$'s.
First limits from the LHC on the mass 
of such a particle are established, excluding 
$m(T)<300 \GeV$ for $m(A_0) = 10 \GeV$
and 
$m(T)<275 \GeV$ for $m(A_0) = 50 \GeV$ 
with 95\% confidence.

\section{Conclusion and Outlook}
With the first 35 \ipb\ of $pp$ collision data collected at $\sqrt{s} = 7 \TeV$ in 2010, 
a suite of top-quark measurements has been presented by ATLAS. 
Some of these measurements have uncertainties that are already 
competitive with uncertainties of theoretical predictions.
For instance, $\stt$ is now measured with an accuracy at the level of $10\%$. 
With the increase of the dataset by two orders
of magnitude the whole spectrum of top physics can be explored at LHC.

\section*{Acknowledgments}
I wish to thank the LHC crew and the ATLAS Collaboration for the excellent data quality,
the many authors and contributors of the top physics analyses shown here, 
Stan Bentvelsen and Wouter Verkerke for 
reading the manuscript and the organizers of these {\em Recontres} for keeping the {\em spirit of Moriond}
alive.  I also gratefully acknowledge the support of the Deutsche 
Forschungsgemeinschaft (DFG) through the Emmy-Noether grant CR-312/1-1.

\end{document}